# Enhancing Science Literacy through Cognitive Conflict-Based Generative Learning Model: An Experimental Study in Physics Learning

## Akmam Akmam[1,*] | Serli Ahzari[1] | Emiliannur Emiliannur[1] | Rio Anshari[1] | David Setiawan[2]

[1]Department of Physics, Faculty of Mathematics and Natural Sciences, Universitas Negeri Padang, Padang, Indonesia
[2]Department of Electrical Engineering, Faculty of Engineering, Universitas Lancang Kuning, Pekanbaru, Indonesia
[*]Corresponding author

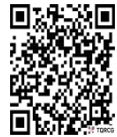

**Abstract:**

This experimental study investigates the effectiveness of the Cognitive Conflict-Based Generative Learning Model (GLBCC) in enhancing science literacy among Indonesian high school physics students. The novelty of this research lies in the innovative integration of cognitive conflict strategies with generative learning principles through a six-stage structured framework, specifically designed to address persistent misconceptions in physics education while systematically developing scientific literacy competencies. The research employed a quasi-experimental pretest-posttest control group design involving 167 Grade XI students from three schools. Students were randomly assigned to experimental groups (n = 83) that received GLBCC instruction and control groups (n = 84) that used the expository learning model. Science literacy was measured using validated instruments assessing scientific knowledge, inquiry processes, and application skills across six key indicators. Statistical analysis using ANOVA with Tukey HSD post-hoc tests revealed significant improvements in science literacy scores for students receiving GLBCC instruction compared to traditional methods (p < 0.001). This study makes a unique contribution to physics education by demonstrating how the deliberate creation of cognitive conflict, combined with authentic real-world physics phenomena, can effectively restructure students' conceptual understanding and enhance their scientific thinking capabilities. Factor analysis identified four critical implementation factors: science literacy development components, learning stages and orientation, motivation and objectives, and knowledge construction processes. The findings provide empirical evidence supporting the integration of cognitive conflict strategies with generative learning approaches in physics education, offering practical implications for educators seeking to enhance students' 21st-century science literacy skills.

**Keywords:** Science Literacy, Generative Learning Model, Cognitive Conflict, Physics Learning

**Introduction:**

Physics education plays a crucial role in developing students' science literacy in the 21st century. Science literacy, defined as the ability to understand, apply, and evaluate scientific concepts, has become increasingly essential for students to navigate the complex global challenges of the 21st century [1]. Effective physics learning requires contextual approaches that actively engage students and enable comprehensive understanding of concepts within real-world environments [2]. However, current physics education faces significant challenges in enhancing students' science literacy, particularly in developing countries.

The Programme for International Student Assessment (PISA) 2018 results revealed concerning gaps in the science literacy performance of Indonesian students. Indonesian students achieved average scores of 371, 379, and 396 in reading, mathematics, and science, respectively, all of which fell substantially below the international averages of 487, 489, and 489 [3]. Similarly, physics learning achievements in Indonesian senior high schools continue to demonstrate suboptimal performance [4], [5], [6]. These findings highlight several critical issues: the implementation of inappropriate learning models that fail to align with student characteristics, ineffective utilization of educational technology [7], [8], and insufficient student interest and motivation in physics subjects [9]. These challenges are further compounded by traditional teaching approaches that emphasize rote memorization over conceptual understanding, limiting students' ability to apply scientific principles to real-world contexts. The lack of hands-on laboratory experiences and interactive learning opportunities has also contributed to students' disconnect from physics concepts, making the subject appear abstract and irrelevant to their daily lives.

Current instructional strategies have proven inadequate in strengthening science literacy within meaningful physics learning contexts [10]. This inadequacy primarily stems from the insufficient contextualization of physics content. Students struggle to connect abstract physics concepts with real-world applications through the cognitive processes of assimilation and accommodation, which are necessary for building a more profound understanding [11]. Consequently, they experience difficulty applying physics knowledge to everyday situations, limiting their development of functional science literacy.

Addressing these educational gaps is critically important for several interconnected reasons. First, strong conceptual understanding serves as the foundation for effective physics problem-solving, enabling students to develop quantitative solutions to qualitative problems [12]. Second, the current digital era demands a generation equipped with robust science literacy to meet emerging technological and global challenges. Third, evidence-based instructional strategies are urgently needed to transform physics education and improve student learning outcomes. The implementation of innovative pedagogical approaches, especially those validated through experimental research, can provide the necessary framework for enhancing science literacy in physics education.

Cognitive conflict-based generative learning model offer a promising solution to address these identified challenges. Cognitive conflict theory posits that when students encounter information contradicting their existing physics concepts, the resulting psychological disequilibrium motivates them to restructure their understanding, leading to conceptual change and deeper learning [13]. This approach proves particularly valuable in physics education, where students commonly hold persistent misconceptions about fundamental concepts. The generative learning model complements this approach by requiring students to actively construct new understanding through connecting prior knowledge with new experiences [14]. The integration of these theoretical frameworks creates powerful learning experiences that systematically challenge students' preconceptions while promoting the construction of meaningful knowledge in physics.

The implementation of the cognitive conflict-based generative learning model utilizes authentic case studies derived from natural phenomena. Through this approach, students actively engage in studying daily physics phenomena and develop science literacy to solve real-world physics problems. The cognitive conflict component deliberately introduces discrepant events or contradictory information to challenge existing physics concepts, while the generative learning component facilitates active knowledge construction through structured learning experiences.

Based on this theoretical framework and the identified educational needs, this experimental study investigates the effectiveness of implementing the cognitive conflict-based generative learning model in enhancing science literacy among senior high school students. This research contributes to the empirical evidence base for innovative physics pedagogy, providing practical insights for educators seeking to enhance students' science literacy. Furthermore, given the global emphasis on science literacy as a fundamental 21st-century skill, the findings of this experimental study have implications extending beyond the Indonesian context, potentially informing evidence-based physics education practices internationally. The present experimental study aims to address the following research questions:

1) How does the Cognitive conflict-based generative learning model influence students' science literacy in senior high school physics learning?

2) What factors influence the implementation of the Cognitive conflict-based generative learning model for achieving students' science literacy in senior high school physics learning?

**Literature Review:**

*A. Cognitive Conflict-Based Generative Learning Model*

The Generative Learning Model with Cognitive Conflict Strategy (GLBCC) is a learning pattern that encourages students to actively engage in the learning process by creating new knowledge through interaction with learning materials [15]. The GLBCC model has garnered attention due to its potential to enhance students' conceptual understanding and higher-order thinking skills. This model is rooted in constructivist theory, which emphasizes the active role of students in constructing their knowledge.

The GLBCC model emphasizes active student involvement in constructing their knowledge. Students are encouraged to assimilate and construct information through interactive learning experiences. It aligns with constructivist principles, where students not only receive information but also participate in the learning process [16]. In the generative learning model, students do not merely receive information passively; instead, they actively construct new understanding by connecting the received information with their previously acquired knowledge [17]. This active construction process enables students to develop a deeper conceptual understanding and engage in meaningful learning experiences.

Furthermore, the generative learning model with cognitive conflict strategy consists of 6 syntaxes: 1) orientation, which functions as the activation of prior knowledge to produce meaningful learning through cognitive processes, 2) cognitive conflict, which aims to stimulate students' curiosity, 3) disclosure, where students are encouraged to consider problem-solving strategies in the form of cognitive conflict, 4) construct, which aims for students to develop conceptual knowledge, 5) application, which aims to encourage students to practice what they have learned, expand their knowledge and skills, and 6) reflective evaluation, which aims to provide feedback on the construction process and the achieved results [14], [15], [18]. The following explanation in Figure 1.

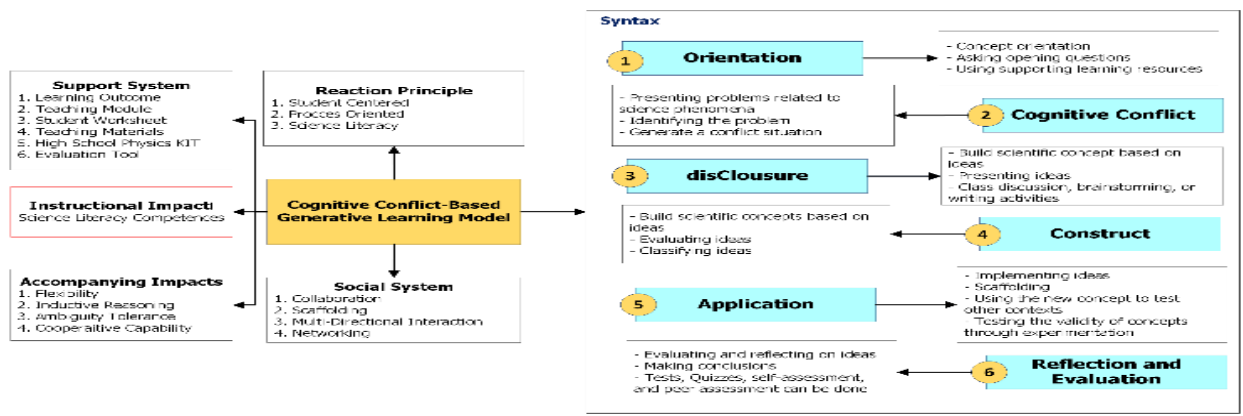

**Fig 1. The Cognitive Conflict-Based Generative Learning Model (GLBCC) Framework**

### B. Ekspository Learning

The expository learning model is a learning strategy that emphasizes the process of delivering learning materials verbally from educators to students, enabling students to master the material optimally. According to recent research, the expository learning strategy is a learning approach that emphasizes the process of delivering material verbally from a teacher to a group of students, with the intention that students can understand the learning material optimally [19], [20], [21]. In the context of its implementation, the expository learning model is a learning approach that emphasizes the process of delivering learning materials verbally from a tutor, educator, or teacher to students or trainees, to enable participants to master the material optimally [22]. The primary characteristic of this model is the dominance of the teacher's role as the primary information provider, with students acting as recipients of information delivered directly to them.

The effectiveness of the expository learning model has been demonstrated in various empirical studies, which show positive impacts on student learning achievement. Research findings show that the implementation of expository learning strategies significantly enhances students' understanding of social studies subjects [20]. Similar findings are also confirmed by other research stating that the expository method is a learning method that emphasizes the process of delivering material verbally from a teacher to a group of students with the intention that students can master the learning material optimally, and has proven effective in improving learning achievement [23]. The main advantage of this model lies in its time efficiency in delivering structured material, as well as the teacher's ability to control the sequence and depth of material presented to students.

The implementation of the expository learning model in modern educational contexts demonstrates high flexibility across various subjects and educational levels. Research shows that the implementation of expository learning models in enhancing students' mathematical learning activities and achievement yields significant results.

Furthermore, the expository learning strategy is an approach that emphasizes the verbal delivery of material by teachers to groups of students, enabling them to master the learning material optimally. It has proven effective in enhancing students' understanding of Indonesian language learning materials [24]. Nevertheless, the implementation of the expository model needs to be integrated with other, more interactive learning methods to create more comprehensive and meaningful learning experiences for students in the modern educational era.

### C. Science Literacy

Science literacy represents a fundamental capability that enables individuals to understand the characteristics of science, analyze how science and technology influence the natural, intellectual, and cultural world, and develop a willingness to engage with and care about scientific issues [3]. In contemporary educational contexts, science literacy underlies most science education curricula with the assumption that citizens must be able to appreciate the relevance and utilize scientific knowledge and practices in various personal and social problems. This capability encompasses not only conceptual understanding but also the skills to engage with science-related issues and scientific ideas as reflective and critical human beings, as science literacy plays an important role in lifelong science learning [25]. The development of science literacy is therefore essential for fostering informed citizenship in modern society.

The development of the significant data era and the massive amount of digital information present significant challenges in the form of scientific misinformation that can influence public decision-making [26]. Science literacy plays a crucial role in raising the next generation of adults committed to climate change mitigation by reducing daily household energy consumption [27]. However, the information paradox phenomenon demonstrates that individuals tend to accept information uncritically, even when it is false, while conversely rejecting information that contradicts their worldview, even when it is true. This reality has made research on science literacy a current scientific agenda and the object of various studies. Therefore, instilling intellectual values such as open-mindedness becomes an essential component in developing science literacy, providing important implications for educational practice [28]. These challenges underscore the pressing need for comprehensive approaches to science literacy education that encompass both cognitive and affective dimensions of learning.

Scientifically literate students must possess the competencies to explain phenomena scientifically, evaluate and design scientific investigations, and scientifically interpret data and evidence. These competencies are elaborated in indicators that include the ability to explain facts, concepts, principles, and laws; present hypotheses, theories, and models; and answer questions related to scientific knowledge or information [29]. The implementation of science literacy in learning aims to develop students' abilities to utilize scientific information to solve problems in daily life and generate beneficial scientific resources. The fundamental issue is the increasing development of utilizing scientific information possessed by students to solve problems in daily life and produce beneficial scientific resources from science literacy [30]. These competencies and indicators serve as crucial benchmarks for evaluating the effectiveness of science literacy programs in educational settings.

Science literacy skills can be grouped into two interrelated main categories. First, the skills to recognize and analyze scientific methods, which include the ability to identify valid scientific arguments, evaluate the validity of sources, assess the use and misuse of scientific information, and understand the elements of research design and draw scientific conclusions. Second, the skills of organizing, analyzing, and interpreting quantitative data and scientific information encompass the ability to create graphical representations of data, read and interpret these representations, solve problems using quantitative skills, including probability and statistics, and draw correct conclusions and predictions based on quantitative data. The integration of these two skill categories is crucial for developing critical and analytical thinking in science, which supports evidence-based decision-making and a better understanding of the natural world. Science literacy not only encompasses understanding of the scientific methods used in research but also the ability to process and interpret data obtained through experiments or scientific studies [31]. The mastery of these interconnected skills enables individuals to become scientifically literate citizens capable of making informed decisions in an increasingly complex scientific and technological world.

**Materials and Methods:**

*D. Design Research*

This research employed a quasi-experimental design with a two-group pretest-posttest design [32], [33]. The design was selected to evaluate the causal relationship between the GLBCC intervention and science literacy outcomes while controlling for potential confounding variables. The study implementation was divided into two measurement phases: pre-research and post-research, conducted across three different schools. The pretest was administered after the preliminary research to assess the initial capabilities of the research subjects. Subsequently, the posttest was conducted after the treatment to evaluate students' scientific literacy abilities following the implementation of the GLBCC model. The comparison between pretest and posttest scores was used to determine the effect of implementing the GLBCC model on students' scientific literacy abilities. The treatment effect results were used to assess the effectiveness of implementing the GLBCC learning model in high school physics learning, as illustrated in Table 1.

**Table 1. Desain *quasi eksperiment* dengan *pretest-posttest control gorup***

| Class | Number | Pre-test | Treatment | Post-test |
|---|---|---|---|---|
| Control | 84 students | Test | EL | Test |
| Experiment | 83 students | Test | GLBCC | Test |

The quasi-experimental design with a pretest-posttest control group involved 84 students in the control group, who received Expository Learning (EL) treatment, and 83 students in the experimental group, who received GLBCC treatment. Both groups underwent pre- and post-assessments to measure the impact of the intervention.

### E. Participants and Setting

The study involved 167 high school students from three different schools during the July-December 2024 academic semester. Schools were selected based on similar demographic characteristics, curriculum alignment, and institutional support for research activities. Students were randomly assigned to experimental groups that received GLBCC treatment (n = 83) and control groups that used expository learning (EL) (n = 84).

Participating schools represented diverse geographical locations to enhance external validity and generalizability of findings. School A included 55 students (27 GLBCC, 28 EL), School B included 55 students (27 GLBCC, 28 EL), and School C included 57 students (29 GLBCC, 28 EL). This distribution ensured balanced representation across different educational contexts.

Inclusion criteria required students to be: (1) enrolled in Grade XI physics learning, (2) present for both pretest and posttest assessments, (3) participants in at least 80% of instructional sessions, and (4) provided informed consent for research participation. Students with documented learning disabilities or those who had taken extended absences were excluded from the analysis to maintain data integrity.

### F. Data Collection and Instruments

Science literacy was measured using validated instruments explicitly developed for physics education contexts. The instrument assessed six key dimensions corresponding to the indicators shown in Table 2.

**Table 2. Indicators of science literacy**

| No | Indicators of Science Literacy |
|---|---|
| 1 | Students can describe methods of scientific inquiry and apply them to investigating, questioning, and solving problems. |
| 2 | Students can describe and carry out experimental procedures. |
| 3 | Students can perform laboratory tasks appropriate to the field. |
| 4 | Students can interpret and communicate scientific information using written, oral and/or graphical means |
| 5 | Students can describe and analyze one or more relationships among science, technology and society and demonstrate an understanding of scientific applications in everyday life |
| 6 | Students can demonstrate logical reasoning in explaining natural phenomena, experimental procedures or outcomes, and/or application of scientific or technological concepts. |

The measurement instrument underwent rigorous validation processes, including expert review, pilot testing, and psychometric analysis. Reliability coefficients exceeded 0.87 for all subscales, indicating adequate internal consistency. Content validity was established through expert panel evaluation, while construct validity was confirmed through factor analysis. The instrument was deemed valid and reliable for research implementation.

### G. Data Analysis Techniques

The data was analyzed using SPSS 25. Data analysis employed multiple statistical procedures to comprehensively address the research objectives. Analysis of variance (ANOVA) was used to compare mean differences between experimental and control groups. Tukey's HSD post-hoc tests provided detailed pairwise comparisons when significant main effects were detected.

Factor analysis was employed to identify key components influencing GLBCC implementation. Kaiser-Meyer-Olkin (KMO) and Bartlett's sphericity tests were used to assess the adequacy of the data for factor analysis. Principal component analysis with varimax rotation extracted underlying factors and determined their relative importance.

**Result:**

### H. Effectiveness of GLBCC Model Implementation in Enhancing Students' Science Literacy

The primary objective of this study was to determine the effectiveness of the Generative Learning Based on Cognitive Conflict (GLBCC) model in enhancing science literacy among high school physics students. Analysis of variance was conducted to examine students' science literacy abilities following the implementation of both the GLBCC model and Expository Learning (EL) model. The results of the variance analysis comparing students' science literacy abilities in the experimental and control classes using the ANOVA test with Tukey HSD statistics are presented in Table 3.

**Table 3. Analysis of variance of students' science literacy abilities in the implementation of GLBCC and EL models with Tukey HSD statistics**

| (I) Class | Schools (J) | Mean Difference (I-J) | Std. Error | Sig. | 95% Confidence Interval | |
|---|---|---|---|---|---|---|
| | | | | | Lower Bound | Upper Bound |
| PreTest School A GLBCC | PreTest School A EL | 1.906 | 1.922 | .998 | -4.420 | 8.234 |
| | PreTest School B GLBCC | -24.565 | 1.851 | .000 | -30.657 | -18.473 |
| | PreTest School B EL | -6.236 | 1.922 | .058 | -12.563 | .0912 |
| | PreTest School C GLBCC | -11.207 | 1.891 | .000 | -17.432 | -4.982 |
| | PreTest School C EL | -8.162 | 1.877 | .001 | -14.340 | -1.984 |
| Post School A GLBCC | Post School A EL | 20.892 | 1.922 | .000 | 14.565 | 27.220 |
| | Post School B GLBCC | 1.893 | 1.851 | .997 | -4.198 | 7.986 |
| | Post School C GLBCC | 7.000 | 1.891 | .132 | .774 | 13.225 |
| PreTest School B GLBCC | PreTest School B EL | 18.329 | 1.813 | .643 | 12.359 | 24.298 |
| | PreTest School C GLBCC | 52.177 | 1.813 | .000 | 46.012 | 58.342 |
| Post School B GLBCC | Post School B EL | 34.177 | 1.813 | .000 | 28.208 | 40.146 |
| | Post School C GLBCC | 5.106 | 1.780 | .158 | -.754 | 10.966 |
| Pretest School C GLBCC | PreTest School D EL | 3.045 | 1.807 | .874 | -2.904 | 8.995 |
| Posttest School C GLBCC | Post School C EL | 15.887 | 1.807 | .000 | 9.937 | 21.837 |

Table 3 presents the comprehensive analysis of variance results for students' science literacy abilities across different schools implementing both GLBCC and EL models. The statistical analysis reveals significant differences between pretest and posttest scores, with the experimental groups consistently demonstrating superior performance. Notably, the Post School A GLBCC group showed a mean difference of 20.892 compared to the Post School A EL group (p < 0.001), indicating substantial improvement in science literacy abilities. Similarly, Post School C GLBCC demonstrated a mean difference of 7.000 compared to baseline measurements (p < 0.132), suggesting meaningful educational gains.

These findings demonstrate that students receiving GLBCC model instruction achieved significantly higher science literacy scores compared to those receiving traditional expository learning at the 5% significance level. The statistical evidence supports the hypothesis that the GLBCC learning model produces significantly positive effects on students' science literacy abilities in high school physics education at the grade XI level. This improvement can be attributed to the model's emphasis on engaging students in systematic development of scientific, logical, and critical thinking processes, which are fundamental competencies for understanding complex physics concepts.

## I. Factor Analysis of GLBCC Model Implementation

The secondary objective was to identify and analyze factors that influence the successful implementation of the GLBCC model in achieving enhanced science literacy outcomes. A comprehensive factor analysis was conducted using 19 initial variables to determine their impact on learning achievement and the development of science literacy. The Kaiser-Meyer-Olkin (KMO) measure and Bartlett's Sphericity test were employed to assess sampling adequacy and data suitability for factor analysis. The result of the KMO and Bartlett test are presented in Table 4.

**Table 4. KMO and Bartlett test of Factors Influencing GLBCC Model on Science Literacy**

| Kaiser-Meyer-Olkin Measure of Sampling Adequacy. | | .637 |
|---|---|---|
| Bartlett's Test of Sphericity | Approx. Chi-Square | 151.018 |
| | df | 78 |
| | Sig. | .000 |
| a. Based on correlations | | |

The initial KMO measure yielded a value of 0.637, indicating adequate sampling sufficiency for factor analysis procedures. The Bartlett's Test of Sphericity produced a chi-square value of 151.018 with 78 degrees of freedom (p < 0.001), confirming strong inter-variable correlations suitable for factor extraction. These preliminary results validated the appropriateness of conducting factor analysis on the proposed variables. Variables that strongly influence the model were identified by continuing with Anti-image Correlation analysis, with the condition that if the MSA value ≥ 0.5, then the factors proposed in the model are suitable for continuation, as shown in Table 5.

**Tabel 51. Anti-image Correlation of GLBCC model influence on physics learning toward high school students' science literacy abilities**

| Variables | 1 | 2 | 3 | 4 | 5 | 6 | 7 | 8 | 9 | 10 | 11 | 12 | 13 |
|---|---|---|---|---|---|---|---|---|---|---|---|---|---|
| GLBCC Implementation Guidelines | .452[a] | -.151 | -.081 | -.109 | -.035 | .047 | -.229 | .146 | -.177 | -.074 | .014 | -.065 | -.080 |
| Learning Objectives | -.151 | .580[a] | .056 | .253 | -.158 | .008 | -.060 | .090 | -.145 | .227 | -.039 | .045 | -.066 |
| Orientation | -.081 | .056 | .533[a] | .195 | -.196 | .085 | -.125 | .002 | -.167 | -.016 | -.014 | .036 | -.024 |
| Stages | -.109 | .253 | .195 | .517[a] | .076 | .066 | .003 | .000 | -.068 | .052 | -.153 | .061 | -.059 |
| Cognitive Conflict | -.035 | -.158 | -.196 | .076 | .793[a] | -.210 | .018 | -.148 | .279 | .119 | .079 | .042 | -.133 |
| Idea Expression | .047 | .008 | .085 | .066 | -.210 | .794[a] | .046 | -.155 | -.057 | -.118 | .016 | .066 | -.503 |
| Construct | -.229 | -.060 | -.125 | .003 | .018 | .046 | .651[a] | -.214 | .112 | .022 | -.151 | .101 | .049 |
| Application | .146 | .090 | .002 | .000 | -.148 | -.155 | -.214 | .801[a] | -.185 | .236 | -.087 | -.118 | -.325 |
| Reflection and Evaluation | -.177 | -.145 | -.167 | -.068 | .279 | -.057 | .112 | -.185 | .342[a] | -.144 | .228 | -.054 | -.044 |
| Learning Motivation | -.074 | .227 | -.016 | .052 | .119 | -.118 | .022 | .236 | -.144 | .523[a] | -.109 | -.090 | -.035 |
| Implementation for Literacy | .014 | -.039 | -.014 | -.153 | .079 | .016 | -.151 | -.087 | .228 | -.109 | .725[a] | -.295 | -.419 |
| Time Adequacy | -.065 | .045 | .036 | .061 | .042 | .066 | .101 | -.118 | -.054 | -.090 | -.295 | .411[a] | .112 |
| Science Literacy | -.080 | -.066 | -.024 | -.059 | -.133 | -.503 | .049 | -.325 | -.044 | -.035 | -.419 | .112 | .746[a] |

Anti-image correlation analysis was subsequently performed to identify variables with sufficient explanatory power for inclusion in the final model. Variables with Measure of Sampling Adequacy (MSA) values below 0.5 were systematically excluded from further analysis. Three variables failed to meet the minimum threshold criteria: GLBCC implementation guidelines (MSA = 0.452), reflection and evaluation guidelines (MSA = 0.342), and time adequacy (MSA = 0.41). After filtering, the KMO and Bartlett test results for the remaining 10 variables are presented in Table 6.

**Table 6. KMO and Bartlett test of factors influencing GLBCC model implementation after filtering**

| KMO and Bartlett's Test | | |
|---|---|---|
| Kaiser-Meyer-Olkin Measure of Sampling Adequacy. | | .772 |
| Bartlett's Test of Sphericity | Approx. Chi-Square | 236.556 |
| | df | 45 |
| | Sig. | .000 |

The KMO MSA measurement result is 0.772. This result indicates that the data meets the requirements for further analysis. The anti-image correlation analysis reveals that the MSA values of the 10 variables proposed in the model exceed 0.5, indicating suitability for continuation. The KMO MSA value of greater than 0.5 indicates a strong partial correlation between each variable that influences science literacy ability in high school Physics learning, as analyzed by the GLBCC model. Each variable can be further predicted and analyzed. The Bartlett Sphericity test results obtained a significant level (p = 0.0000). The value (p < 0.0001) indicates a significant correlation between each variable influencing science literacy ability in physics learning and the GLBCC model.

The following process involves extracting a collection of factor components (variables) into several factors using a communality analysis. The communalities analysis results yielded determination coefficients for each component (variable) against the formed factors, ranging from 0.522 to 0.879. The communalities coefficient values of all components are greater than 0.05. This result indicates that the analysis process can be continued by extracting variables with multivariate methods that transform correlated original factor components (variables) into new uncorrelated factors. The analysis results of the total variance of each variable show that ten variables can form four new factors with total eigenvalues of 3.299, 1.555, 1.068, and 1.0047, respectively.

**Table 7. Principal Component Analysis Results with Extraction Method of GLBCC Model Implementation**

| Component | Initial Eigenvalues | | | Rotation Sums of Squared Loadings | | | | | |
|---|---|---|---|---|---|---|---|---|---|
| | Total | % of Variance | % Cumulative | % of Variance | % Cumulative | | Total | % of Variance | % Cumulative |
| 1 | **3.299** | 32.990 | 32.990 | 32.990 | 32.990 | | 3.101 | 31.006 | 31.006 |
| 2 | **1.555** | 15.547 | 48.537 | 15.547 | 48.537 | | 1.350 | 13.499 | 44.505 |
| 3 | **1.068** | 10.678 | 59.215 | 10.678 | 59.215 | | 1.334 | 13.337 | 57.842 |
| 4 | **1.004** | 10.039 | 69.254 | 10.039 | 69.254 | | 1.141 | 11.412 | 69.254 |
| 5 | .823 | 8.231 | 77.485 | | | | | | |
| 6 | .666 | 6.662 | 84.148 | | | | | | |
| 7 | .567 | 5.674 | 89.822 | | | | | | |
| 8 | .469 | 4.686 | 94.508 | | | | | | |
| 9 | .353 | 3.529 | 98.037 | | | | | | |
| 10 | .196 | 1.963 | 100.000 | | | | | | |

The final stage of analysis involves rotating the component matrix using the varimax method, which is a component extraction technique that maximizes weighting factors through the Kaiser normalization method. The loading factor values show the magnitude of correlation between the formed factors and the constituent variables. The component matrix rotation results are presented in Table 5, and then the data are grouped. The grouping of variables into each formed factor is based on the loading factor values (correlation values) of each variable. The loading factor value used is $(0.7 \le r < 0.9)$, which means there is a strong relationship between variables and the formed factors. The determination of indicators in forming factors is evident from the comparison of the magnitude of correlation values in each row. The grouping results of the five formed factors are presented in Table 8.

**Table 8. Rotated Component Matrix Factor Analysis Affecting GLBCC Model Implementation**

| Variable | Component | | | |
|---|---|---|---|---|
| | 1 | 2 | 3 | 4 |
| Science Literacy | **.917** | -.006 | -.021 | .064 |
| Idea Expression | **.868** | .093 | -.008 | -.092 |
| Application | **.778** | -.049 | -.182 | .194 |
| Implementation for Literacy | **.697** | -.221 | .140 | .243 |
| Cognitive Conflict | **.607** | .330 | -.326 | .033 |
| Stages | .055 | **-.797** | .141 | .125 |
| Orientation | .034 | **.632** | .151 | .520 |
| Learning Motivation | -.003 | .105 | **.866** | -.104 |
| Learning Objectives | .126 | .364 | **-.601** | -.009 |
| Construct | .160 | -.051 | -.143 | **.857** |

Table 8 shows that 10 components (variables) influencing the learning process are grouped into four closely related factors. The first factor consists of Science Literacy, Idea Expression, Application, and Implementation for Literacy, as well as Cognitive Conflict. The second factor consists of two variables: learning stages and orientation. The third factor consists of two variables: Stages and Orientation in the application of the GLBCC model. The fourth factor has only one variable: Construct. As the final step in determining factors, analyzing the Component Transformation is crucial. The Component Transformation Matrix helps see the magnitude of correlation values from the formed factors. The correlation values in Component Transformation are shown in Table 9.

**Table 9. Component Transformation Matrix Test Results for Factors Affecting GLBCC Model Implementation**

| Component Transformation Matrix | | | | |
|---|---|---|---|---|
| Component | 1 | 2 | 3 | 4 |
| 1 | .949 | .089 | -.199 | .227 |
| 2 | .226 | -.750 | .609 | -.120 |
| 3 | .003 | .530 | .734 | .424 |
| 4 | -.218 | -.385 | -.223 | .869 |
| Extraction Method: Principal Component Analysis. | | | | |
| Rotation Method: Varimax with Kaiser Normalization. | | | | |

Table 9 shows that the correlation values of each formed factor are greater than 0.5 (0.949 for factor 1, 0.750 for factor 2, 0.734 for factor 3, and 0.869 for factor 4). The negative sign only indicates the direction of correlation, but these numbers can still summarize the existing variables. Correlation coefficients greater than 0.5 indicate that the four formed factors are suitable for summarizing the 10 variables analyzed.

**Discussion:**

The ANOVA test results show that the class receiving the GLBCC model learning achieved higher science literacy ability scores compared to the class receiving the Expository Learning (EL) model at a 5% significance level. The data indicates that the GLBCC learning model has a significantly positive impact on students' science literacy abilities in high school Physics learning in Grade XI. The superior performance occurs because students who receive the GLBCC learning model are always involved in developing scientific, logical, and critical thinking [34], [35]. Students require cognitive abilities to comprehend physics material.

The statistical analysis results in Table 3 show that the science literacy ability of students who received the GLBCC learning model is higher compared to the average science literacy ability scores of students who received the EL learning model. This suggests that the science literacy embedded in students' minds persist in their thinking [35]. Learning is a process of behavioral change through systematic activities in building knowledge to achieve goals, improve understanding, attitudes, behaviors, skills, and habits. The behavioral change process occurs due to interactions between stimulus and response. A person is considered to be in the behaviorist learning concept when behavioral changes occur in a positive direction and demonstrates flexibility in their behavior. Behaviors that emerge in response to stimuli do not appear spontaneously but must go through a thinking process (cognitive process). Learning is a cognitive process toward repeatedly presented stimuli with reinforcement. Learning must involve activities that engage complex thinking processes. This type of thinking pattern encourages students to think creatively to solve problems they face.

The data analysis results for Grade XI Physics learning show that there is no significant difference in science literacy abilities between experimental and control class students at the beginning of Physics learning in high school. The science literacy abilities of experimental and control classes differ significantly after both receive different treatments. The experimental class learning uses the generative GLBCC learning model, while the control class uses the guided inquiry learning model with probing question strategies. Generally, the science literacy ability of the experimental class is higher than that of the control class. These results indicate that the GLBCC model has a positive effect on students' science literacy in Grade XI high school Physics subjects. Therefore, the GLBCC model is effective for high school Physics learning. This occurs because students who receive the GLBCC learning model are always involved in developing scientific, logical, and critical thinking [34], [35]. The GLBCC model has facilitated science literacy, a global trend in education that is essential for the development of science and technology [30], [36].

Through the GLBCC model, students actively understand data and integrate it with previous knowledge, enabling them to apply it to new situations [37]. The GLBCC model can help address fundamental problems associated with the increasing use of scientific information, enabling students to solve problems in daily life and produce beneficial scientific sources from science literacy, a significant issue in its development [38]. Through the GLBCC model, students are encouraged to strive to understand, interpret, and apply science concepts and information provided. GLBCC learning activities that facilitate students will ensure they gain lasting experiences, making learning practical and effective.

In the orientation stage, students are already required to think more specifically and diversely. Students are also required to consider more than one idea, which is an important initial step toward success in learning [39]. When students encounter scientific problems that contain cognitive conflict, they will imagine and conceptualize problem-solving solutions, thereby contributing to filling gaps in their prior knowledge [40]. This shows that the orientation syntax can improve fluency and flexibility indicators in students [41]. Students will then be given problems according to the material.

Anomalies in scientific phenomena given to students are implemented in the cognitive conflict syntax. Cognitive conflict contains important information that helps students recognize misconceptions by providing stimuli that function as catalysts for policy change [42]. The integration of augmented reality technology with cognitive conflict models has been shown to enhance students' understanding of complex physics concepts by creating immersive learning experiences that challenge preconceived notions [43]. Stimuli provided in the form of problems encountered in daily life will help students understand these problems more easily [44]. This shows that the cognitive conflict syntax can improve flexibility indicators in students [45]. Educators will then guide students to avoid having different concepts. Knowledge construction is carried out in the construct syntax. The construct syntax involves educators constructing students' knowledge. Educators serve as facilitators who guide students in constructing knowledge quickly and efficiently [46]. Knowledge construction helps students understand concepts.

Application activities in conducting practicum are activities that help students discover concepts through experiments. The implementation of virtual reality in physics education has demonstrated significant improvements in creative thinking skills and self-efficacy, particularly in understanding complex topics like rotational dynamics [47]. Research has also shown that combining augmented reality with cognitive conflict models and STEM integration can effectively support the development of scientific literacy, particularly in teaching fundamental physics concepts such as Newton's Universal Law of Gravitation [48]. Construction results will produce new, original knowledge from the students themselves. This shows that the construct syntax can improve originality indicators in students [41]. Students' constructed knowledge will then be used in problem-solving. The application syntax requires students to apply what they have learned to build knowledge.

The application syntax is implemented by providing problems. Students are required to be able to answer problems independently, provide multiple alternative answers, and offer original responses. This shows that the application syntax can improve fluency, flexibility, and originality indicators in students [49]. Problems that students have solved are then given feedback. Educators provide feedback in the reflection evaluation syntax. Providing feedback to students can improve their scientific literacy and cognitive abilities [50], [51]. Educators instruct students to evaluate the strengths and weaknesses of the information they receive. Educators also direct students to reflect on and conclude their learning. This enables students to create something new and unique that differs from previous works.

**Conclusion:**

This experimental study provides compelling evidence addressing the research questions regarding the influence of the Cognitive Conflict-Based Generative Learning Model (GLBCC) on students' science literacy in senior high school physics learning. The research demonstrates that the GLBCC model significantly enhances science literacy compared to the expository learning model, with students achieving substantially higher scores in scientific thinking, logical reasoning, and critical analysis capabilities. The model's six-stage syntax—Orientation, Cognitive Conflict, Disclosure, Construct, Application, and Reflection-Evaluation—creates a systematic framework for cognitive restructuring and knowledge construction that effectively addresses the challenges of physics education. Through progressive engagement with authentic physics problems, systematic resolution of cognitive conflicts, collaborative knowledge building, and metacognitive reflection, the GLBCC model proves to be a powerful instructional approach for developing students' scientific literacy skills.

Regarding the factors influencing GLBCC implementation, the factor analysis revealed four critical dimensions essential for successful model deployment: science literacy development components, learning stages and orientation, motivation and objectives, and knowledge construction processes. These findings suggest that effective implementation necessitates careful attention to the introduction of systematic cognitive conflict, structured generative learning activities, and appropriate instructional scaffolding that support students' conceptual development. The study's implications extend to educational practice through the need for comprehensive professional development programs, curriculum reforms that incorporate authentic problem-solving contexts, and assessment strategies that evaluate higher-order thinking skills. Future research should investigate the effects of the cognitive conflict-based generative learning model on other physics topics and its influence on other variables that students require in the 21st century. These findings support the development of evidence-based physics learning models that can effectively enhance students' scientific literacy, addressing complex 21st-century challenges.

**Conflict of Interest:**

The authors declare no conflict of interest.

**Author Contributions**

A.A. conceptualized the research framework, designed the experimental methodology, supervised data collection and analysis, and led the manuscript writing process. S.A. conducted the literature review, participated in data collection across multiple schools, performed preliminary statistical analyses, and contributed to the methodology and results sections. E.E. developed and validated the science literacy assessment instrument and contributed to the materials and methods section. R.A. implemented the GLBCC intervention protocols, conducted factor analysis procedures, assisted with statistical interpretations, and contributed to the discussion and conclusion sections. D.S. reviewed statistical analyses for accuracy, contributed to the theoretical framework development, and participated in manuscript revision and editing. All authors reviewed and approved the final manuscript version, contributed to critical revisions, and agreed to be accountable for all aspects of the work's accuracy and integrity.

**Funding**

This research was funded by PNBP UNP with Contract Number 1649/UN.35.15/LT/2024.

**Acknowledgment**

Thank you to Universitas Negeri Padang for facilitating and funding the research. Funding from PNBP UNP with Contract Number 1649/UN.35.15/LT/2024.

**References:**

1. A. Mustakim, S. Jumini, and F. Firdaus, "Pengaruh Penggunaan Modul Pembelajaran Fisika dengan Pendekatan Saintific Berbasis Riset untuk Meningkatkan Literasi Sains Siswa Kelas VIII di SMP Takhassus Al-Qur'an 2 Dero Duwur, di Wonosobo Tahun Ajaran 2018/2019," *Pros. Semin. Nas. Pendidik. Fis. FITK UNSIQ*, vol. 2, no. 1, pp. 217–226, 2020.

2. E. Petričević, "The Contextual and Individual Determinants of Engagement in Learning Physics," *Psihol. Teme*, vol. 31, no. 2, pp. 383–402, 2022, doi: 10.31820/pt.31.2.9.

3. OECD, *PISA 2018 Results: What Students Know and Can Do (Volume I)*, vol. I. Paris: OECD Publishing, 2019. doi: 10.1787/5f07c754-en.

4. A. Amsikan, "Application of Project Based Learning Model to Increase Students Physics Learning Outcomes and Science Process Skills," *Paedagogia*, vol. 25, no. 1, p. 1, 2022, doi: 10.20961/paedagogia.v25i1.58989.

5. S. U. S. Supardi, L. Leonard, H. Suhendri, and R. Rismurdiyati, "Pengaruh Media Pembelajaran dan Minat Belajar Terhadap Hasil Belajar Fisika," *Form. J. Ilm. Pendidik. MIPA*, vol. 2, no. 1, pp. 71–81, 2015, doi: 10.30998/formatif.v2i1.86.

6. E. Haataja, M. Dindar, J. Malmberg, and S. Järvelä, "Individuals in a group: Metacognitive and regulatory predictors of learning achievement in collaborative learning," *Learn. Individ. Differ.*, vol. 96, no.

March, 2022, doi: 10.1016/j.lindif.2022.102146.

7. L. Linkola, C. J. Andrews, and T. Schuetze, "An agent based model of household water use," *Water (Switzerland)*, vol. 5, no. 3, pp. 1082–1100, 2013, doi: 10.3390/w5031082.

8. M. Jenkins and J. D. Walker, "COVID-19 Practices in Special Education: Stakeholder Perceptions and Implications for Teacher Preparation," *Teach. Educ. J.*, vol. 14, pp. 83–105, 2021.

9. A. A. Hero Yawo, "The Influence of Virtual Physics Laboratory on Senior High School Form one Physics Students Performance and Cognitive Achievement at Bishop Herman College, Kpando, Volta Region- Ghana," *Int. J. Eng. Appl. Sci.*, vol. 7, no. 9, 2020, doi: 10.31873/ijeas.7.09.17.

10. Y. Yan Lu, H. Shyang Lin, F. Lai Lin, and Z. R Hong, "Exploring the Effectiveness of a Scientific Inquiry Creative Workshop in Promoting Senior and Vocational High School Students' Scientific Inquiry Self-efficacy," *J. Res. Educ. Sci.*, vol. 67, no. 4, pp. 177–219, 2022, doi: https://doi.org/10.6209/JORIES.202212_67(4).0006.

11. R. Dolfing, G. Prins, A. M. W. Bulte, A. Pilot, and J. D. Vermunt, "Strategies to Support Teachers Professional Development Regarding Sense-Making in Context-Based Science Curricula," *Science Education*, vol. 105, no. 1. pp. 127–165, 2021. [Online]. Available: wileyonlinelibrary.com/journal/sce

12. Q. X. Ryan, D. Agunos, S. Franklin, M. Gomez-Bera, and E. C. Sayre, "Question characteristics and students' epistemic framing," in *Physics Education Research Conference Proceedings*, 2020, pp. 442–447. doi: 10.1119/perc.2020.pr.Ryan.

13. F. Mufit, F. Festiyed, A. Fauzan, and L. Lufri, "Impact of Learning Model Based on Cognitive Conflict toward Student's Conceptual Understanding," *IOP Conf. Ser. Mater. Sci. Eng.*, vol. 335, no. 1, 2018, doi:

10.1088/1757-899X/335/1/012072.

14. A. Akmam, R. Hidayat, F. Mufit, R. Anshari, and N. Jalinus, "Effect of Generative Learning Models Based on Cognitive Conflict on Students' Creative Thinking Processes Based on Metacognitive," *J. Phys. Conf. Ser.*, vol. 2582, no. 1, p. 012058, Sep. 2023, doi: 10.1088/1742-6596/2582/1/012058.

15. A. Akmam, R. Hidayat, F. Mufit, N. Jalinus, and A. Amran, "Need Analysis to Develop a Generative Learning Model with a Cognitive Conflict Strategy Oriented to Creative Thinking in the Computational Physics Course," *J. Phys. Conf. Ser.*, vol. 2309, no. 1, 2022, doi: 10.1088/1742-6596/2309/1/012095.

16. A. Akmam, R. Hidayat, F. Mufit, R. Anshari, and N. Jalinus, "Effect of Generative Learning Models Based on Cognitive Conflict on Students ' Creative Thinking Processes Based on Metacognitive," in *Journal of Physics: Conference Series*, 2023, p. 012058. doi: 10.1088/1742-6596/2582/1/012058.

17. H. W. Lee, K. Y. Lim, and B. L. Grabowski, "Improving self-regulation, learning strategy use, and achievement with metacognitive feedback.," *Educ. Technol. Res. Dev.*, vol. 58, no. 6, pp. 629–648, 2010, doi: 10.1007/s11423-010-9153-6.

18. A. Akmam, R. Afrizon, I. Koto, D. Setiawan, R. Hidayat, and F. Novitra, "Integration of Conflict in Generative Learning Model to Enhancing Students' Creative Thinking Skills," *Eurasia J. Math. Sci. Technol. Educ.*, vol. 20, no. 9, p. em2504, Sep. 2024, doi: 10.29333/ejmste/15026.

19. J. Jumaisa, "Model Pilihan Pembelajaran, Inquiry atau Expository?," *J. Ilm. Mandala Educ.*, vol. 6, no. 2, pp. 339–348, 2020, doi: 10.58258/jime.v6i2.1441.

20. A. K. Hasan, M. Athila, M. Bertuanda, S. Sapriya, and W. Wilodati, "Relevance of Using Expository Learning Strategies in Teaching And Learning Activities in

Schools," *Prog. Pendidik.*, 2025, [Online]. Available: https://api.semanticscholar.org/CorpusID:276066040

21. W. N. Nasution, "Expository Learning Strategy: Definition, Goal, Profit and Procedure," 2020, [Online]. Available: https://api.semanticscholar.org/CorpusID:219630229

22. W. O. Eli, "Penerapan Model Pembelajaran Ekspositori Dalam Meningkatkan Hasil Belajar IPS Siswa Kelas VIII MTs Negeri 4 Buton Selatan," *J. Akad. Pendidik. Ekon.*, vol. 6, no. 1, pp. 51–66, 2019, [Online]. Available: http://refcale.uleam.edu.ec/index.php/enrevista/article/view/1225

23. A. K. Peterson, C. B. Fox, and M. Israelsen, "A systematic review of academic discourse interventions for school-aged children with language-related learning disabilities," *Lang. Speech. Hear. Serv. Sch.*, vol. 51, no. 3, pp. 866–881, 2020, doi: 10.1044/2020_LSHSS-19-00039.

24. E. Irmayanti, B. Surindra, E. Prastyaningtyas, and T. Ayatik, "Penerapan Model Pembelajaran Ekspositori Untuk Meningkatkan Motivasi, Keaktifan, Kemampuan Memecahkan Masalah, Kolaborasi, dan Hasil Belajar Siswa Dengan Pendekatan Saintifik Berbasis Lesson Study," no. 6, pp. 165–172, 2019, doi: https://doi.org/10.29407/E.V6I2.13754.

25. D. Fortus, J. Lin, K. Neumann, and T. D. Sadler, "The role of affect in science literacy for all," *Int. J. Sci. Educ.*, vol. 44, no. 4, pp. 535–555, 2022, doi: 10.1080/09500693.2022.2036384.

26. T. S. Sheromova, A. N. Khuziakhmetov, V. A. Kazinets, Z. M. Sizova, S. I. Buslaev, and E. A. Borodianskaia, "Learning styles and development of cognitive skills in mathematics learning," *Eurasia J. Math. Sci. Technol. Educ.*, vol. 16, no. 11, 2020, doi: 10.29333/EJMSTE/8538.

27. J. J. B. R. Aruta, "Science literacy promotes

energy conservation behaviors in Filipino youth via climate change knowledge efficacy: Evidence from PISA 2018," *Aust. J. Environ. Educ.*, vol. 39, no. 1, pp. 55–66, 2023, doi: 10.1017/aee.2022.10.

28. A. J. Sharon and A. Baram-Tsabari, "Can science literacy help individuals identify misinformation in everyday life?," *Sci. Educ.*, vol. 104, no. 5, pp. 873–894, 2020, doi: 10.1002/sce.21581.

29. S. Fayanto, S. Sulthoni, A. Wedi, A. Takda, and M. Fadilah, "Exploration of Integrated Science-Physics Textbooks Based on Science Literacy Indicators: A Case Study in Kendari City Indonesia," *Anatol. J. Educ.*, vol. 8, no. 1, pp. 159–172, 2023, doi: 10.29333/aje.2023.8111a.

30. F. Fakhriyah, S. Masfuah, M. Roysa, A. Rusilowati, and E. S. Rahayu, "Student's science literacy in the aspect of content science?," *J. Pendidik. IPA Indones.*, vol. 6, no. 1, pp. 81–87, 2017, doi: 10.15294/jpii.v6i1.7245.

31. C. Gormally, P. Brickman, and M. Lut, "Developing a test of scientific literacy skills (TOSLS): Measuring undergraduates' evaluation of scientific information and arguments," *CBE Life Sci. Educ.*, vol. 11, no. 4, pp. 364–377, 2012, doi: 10.1187/cbe.12-03-0026.

32. J. Creswell, *Research Design: Qualitative, Quantitative and Mixed Methods Approaches*. Pearson Education Inc., 2015.

33. Sugiyono, *Metode Penelitian Kuantitatif, Kualitatif, dan R&D*. Bandung: Alfabeta, 2015.

34. Y. Zhang, *Assessing Literacy in a Digital World: Validating a ScenarioBased Reading-to-Write Assessmen*. Springer Nature, 2022.

35. K. Vlasenko *et al.*, "The Criteria of Usability Design for Educational Online Courses," no. January, pp. 461–470, 2022, doi: 10.5220/0010925200003364.

36. A. Akkam, F. Mufit, R. Hidayat, and R. Anshari, "Pengembangan Model Pembelajaran Generatif Berstrategi Konflik Kognitif Berorientasi Berpikir Kreatif Mahasiswa Pada Mata Kuliah Komputasi Fisika." p. HIBAH UNP (PD), 2022.

37. L. Fiorella and R. E. Mayer, "Eight Ways to Promote Generative Learning," *Educational Psychology Review*, vol. 28, no. 4. 2016. doi: 10.1007/s10648-015-9348-9.

38. F. Mumtaz, S. Sjaifuddin, and A. Nestiadi, "The effect of the generative learning model on the student critical thinking ability in environmental conservation topic," *J. Pijar Mipa*, vol. 18, no. 4, pp. 479–485, 2023, doi: 10.29303/jpm.v18i4.5152.

39. E. Sandoval-Lucero, K. Antony, and W. Hepworth, "Co-Curricular Learning and Assessment in New Student Orientation at a Community College," *Creat. Educ.*, vol. 08, no. 10, pp. 1638–1655, 2017, doi: 10.4236/ce.2017.810111.

40. M. Fleer, "The genesis of design: learning about design, learning through design to learning design in play," *Int. J. Technol. Des. Educ.*, vol. 32, no. 3, pp. 1441–1468, 2022, doi: 10.1007/s10798-021-09670-w.

41. I. N. Suardana, K. Selamet, A. A. I. A. R. Sudiatmika, P. Sarini, and N. L. P. L. Devi, "Guided inquiry learning model effectiveness in improving students' creative thinking skills in science learning," *J. Phys. Conf. Ser.*, vol. 1317, no. 1, 2019, doi: 10.1088/1742-6596/1317/1/012215.

42. E. B. Mandinach and K. Schildkamp, "Misconceptions about data-based decision making in education: An exploration of the literature," *Stud. Educ. Eval.*, vol. 69, no. January 2020, p. 100842, 2021, doi: 10.1016/j.stueduc.2020.100842.

43. F. Mufit and M. Dhanil, "Effectiveness of Augmented Reality with Cognitive Conflict Model to Improve Scientific Literacy of Static Fluid Material," *Int. J. Inf. Educ. Technol.*, vol. 14, no. 9, pp. 1199–1207, 2024, doi: 10.18178/ijiet.2024.14.9.2149.

44. S. Bektiarso, D. R. Dewi, and Subiki, "Effect of problem based learning models with 3D

thinking maps on creative thinking abilities and physics learning outcomes in high school," in *Journal of Physics: Conference Series*, 2021. doi: 10.1088/1742-6596/1832/1/012027.

45. J. Rokhmat, I. W. Gunada, S. Ayub, Hikmawati, and T. Wulandari, "The use of causalitic learning model to encourage abilities of problem solving and creative thinking in momentum and impulse," *J. Phys. Conf. Ser.*, vol. 2165, no. 1, 2022, doi: 10.1088/1742-6596/2165/1/012052.

46. A. Hinck and J. Tighe, "From the other side of the desk: students' discourses of teaching and learning," *Commun. Educ.*, vol. 69, no. 1, pp. 1–18, 2020, doi: 10.1080/03634523.2019.1657157.

47. Yulkifli *et al.*, "The Impact of Virtual Reality on Creative Thinking Skills and Self-Efficacy in Learning Rotational Dynamics," *Int. J. Inf. Educ. Technol.*, vol. 15, no. 6, pp. 1302–1311, 2025, doi: 10.18178/ijiet.2025.15.6.2332.

48. F. Mufit, W. S. Dewi, S. Riyasni, and M. Dhanil, "Augmented Reality with A Cognitive Conflict Model and STEM Integration on Newton's Universal Law of Gravitation: Does Practicing Practical Learning Support Scientific Literacy?," *Int. J. Inf. Educ. Technol.*, vol. 15, no. 2, pp. 255–271, 2025, doi: 10.18178/ijiet.2025.15.2.2239.

49. N. Fauziah, A. Hakim, and Y. Handayani, "Meningkatkan Literasi Sains Peserta Didik Melalui Pembelajaran Berbasis Masalah Berorientasi Green Chemistry Pada Materi Laju Reaksi," *J. Pijar Mipa*, vol. 14, no. 2, pp. 31–35, 2019, doi: 10.29303/jpm.v14i2.1203.

50. H. Dominguez *et al.*, "Learning to Transform, Transforming to Learn: Children's Creative Thinking with Fractions," *J. Humanist. Math.*, vol. 10, no. 2, pp. 76–101, 2020, doi: 10.5642/jhummath.202002.06.

51. [G. G. Calvo and L. M. Álvarez, "Embodied teaching journals as an instrument for reflection and self-evaluation during the teaching practicum," *Estud. Pedagog.*, vol. 44, no. 2, pp. 185–204, 2018, doi: 10.4067/S0718-07052018000200185.